\documentstyle[aps,prl,multicol]{revtex}

\begin{document}
\input{epsf}
\draft
\tighten

\preprint{}
\title{
Metal-Insulator Transition in a Disordered Two-dimensional Electron 
Gas in GaAs-AlGaAs at Zero Magnetic Field
}
\author{E. Ribeiro\cite{evaldo}, R. J\"{a}ggi, T. Heinzel, and K. Ensslin}
\address{Solid State Physics Laboratory, ETH Z\"{u}rich, 8093 
Z\"{u}rich,  Switzerland\\
}
\author{G.  Medeiros-Ribeiro and P. M. Petroff}
\address{Materials Department, University of California, Santa 
Barbara, Ca 93106, USA\\
}
\date{\today}
\maketitle
\begin{abstract}

A metal-insulator transition in two-dimensional electron gases at 
$B=0 $
is found in Ga[Al]As 
heterostructures, where a  high density of 
self-assembled InAs quantum dots is incorporated
just 3 nm below the heterointerface. The transition 
occurs at resistances around 
h/e$^2 $and critical carrier densities of $1.2 \cdot 10^{11}cm^{- 
2}$. Effects of electron-electron
interactions are expected to be rather weak in our samples, while 
disorder plays a crucial role.
\end{abstract}

\begin{multicols} {2}
\narrowtext

The metal-insulator transition (MIT) is one of the central questions 
in the understanding 
of two-dimensional systems\cite{Sondhi97}. Theoretically it has been shown 
that a two-dimensional 
system without interactions is expected to behave as an 
insulator\cite{Abrahams79}. Recent experiments on silicon 
metal-oxide-semiconductor field-effect transistors (MOSFETs) with low 
disorder\cite{Kravchenko94,Kravchenko95,Simonian97,Popovic97}
have clearly demonstrated a MIT at zero magnetic field 
(B=0). Meanwhile, metallic phases
have also been found with holes in SiGe quantum wells\cite{Lam97} and 
 holes in Ga[Al]As 
heterostructures\cite{Simmons98,Hanein98}.
Scaling theory has been crucial 
 in order to characterize the metallic and the insulating states\onlinecite{Sondhi97,Abrahams79}, 
 even in the presence of 
 interactions\cite{Finkelstein83,Dobro97,Castellani98}. The resistivity has been found 
 to scale with temperature as well as 
 with electric field,  in agreement with 
 theoretical considerations\onlinecite{Sondhi97},\cite{Song98}. 
 By studying both scalings, the dynamic exponent z and the 
 correlation length exponent $\nu $ can be obtained 
independently\cite{Kravchenko96}.
In general, samples in 
which a MIT at B=0 was experimentally observed, were optimized towards 
low disorder and 
large electron-electron interaction, i.e. clean samples 
with large effective 
carrier masses and low carrier densities have been used. 

Here, we report on the observation of 
a MIT in a disordered 
two-dimensional electron gas (2DEG) in a Ga[Al]As heterostructure, both as a 
function of temperature and 
electric field at B=0. The MIT in 
our samples is observed at a 
critical carrier densities of $N_{C} = 1.2 \cdot 10^{11} cm^{-2}$, a 
mobility of $2000  cm^{2}/Vs $ and 
at resistances of the order of $h/e^{2}$. The parameters of our 
samples are very different to the ones used in previous work in two respects. First, 
the electron-electron 
interaction energy $E_{ee}$ is comparable to the kinetic energy 
$E_{F}$ of the 
electrons. In our samples, the ratio 
$E_{ee}/E_{F}=(e^{2}/\varepsilon_{0}h^{2}\cdot(m^{*}/\varepsilon  N_{S}^{1/2})$, is 
only 0.9 at $N_{C}$ ($\varepsilon$ denotes the dielectric constant of GaAs, i.e. 
$\varepsilon_{GaAs} =12.8$, and the effective electron mass in GaAs is 
$m^{*}=0.067 m_{0},$ where $m_{0}$ is the free electron mass)
this ratio is $\geq  10$ in the samples studied in 
Refs. \onlinecite{Kravchenko94,Kravchenko95,Simonian97,Popovic97,Lam97,Simmons98,Hanein98,Kravchenko96}. 
Second, our samples are highly disordered. The Drude scattering 
time $\tau_{D}$ at $N_{C}$ is $\tau_{D}=0.08 ps$, almost 2 orders of magnitude below the scattering times in the 
experiments
 cited above. It is also two orders of magnitude below the typical 
dephasing time $\tau_{\phi }\approx 7 ps$ in 
 our samples. The high disorder is generated by a layer of InAs 
self-assembled quantum dots (SAQD) located 
 at the site of the electron gas. The 
details of the scattering centers, 
 namely attractive InAs dots filled with electrons, are rather 
different from conventional disorder 
 predominately originating from residual doping atoms. 
 We observe the 
 MIT at B=0 only in samples with very high dot densities of about 
$5\cdot 10^{10} cm^{-2}.$
Scaling behavior in electric field is found. Scaling in temperature is difficult to confirm because 
of a  poorly defined fix point in the temperature dependence. 
Nevertheless, we can estimate z and 
$\nu$ and find values in the same range as those reported in 
Ref.\onlinecite{Kravchenko96}.
The layout of the paper is as follows: 
first, we will describe the samples and the experimental 
setup. We proceed by 
discussing the temperature dependence of the MIT, followed by its electric 
field dependence. We speculate on 
a possible explanation and conclude with 
a summary .

The details of our samples are described in\cite{Ribeiro98}, but 
the crucial parameters are given below. A layer of InAs SAQDs is embedded in the GaAs buffer 
layer 
3 nm from the $Al_{x}Ga_{1-x}As/GaAs$ interface that defines a 2DEG 
(upper right inset  in Fig. 1). 
The substrate was not rotated during growth of the dot layer, leading 
of a gradient in dot density 
across the wafer from $3\cdot10^{9}$ dots cm$^{-2}$ up to $5\cdot 10^{10}$ 
dots cm$^{-2}$. From TEM studies and transport measurements, the dot 
density is known within a factor of 2 for each part of the wafer. A 
$\delta$-doping (n-type with Si) layer 
within the $Al_{x}Ga_{1-x}As$ 30 nm from the interface provides the carriers for the 2DEG. The 
MIT at B=0 was observed for the 
 sample with the highest density of dots, namely sample 7 from\onlinecite{Ribeiro98}. 
 To ensure a homogeneous dot density, the 
sample size was kept 
sufficiently small. The width of the Hall geometries is 
$20\mu m$, while the voltage probes used to 
measure $\rho_{xx}$ are separated by $40\mu m.$ Gate electrodes are used to change the 
carrier density. The samples are immersed in the mixing chamber of a 
$^{3}He/^{4}He$ dilution refrigerator with 
a base temperature of $60 mK$. 
 Resistivites were measured at low frequencies ($13 Hz$) in a four-probe 
configuration at a current level of 
1nA.  For the electric field 
scaling, the DC bias current was swept and the voltage drop was 
measured.

\begin{figure}
\centerline{\epsfxsize=3.2in\epsfbox{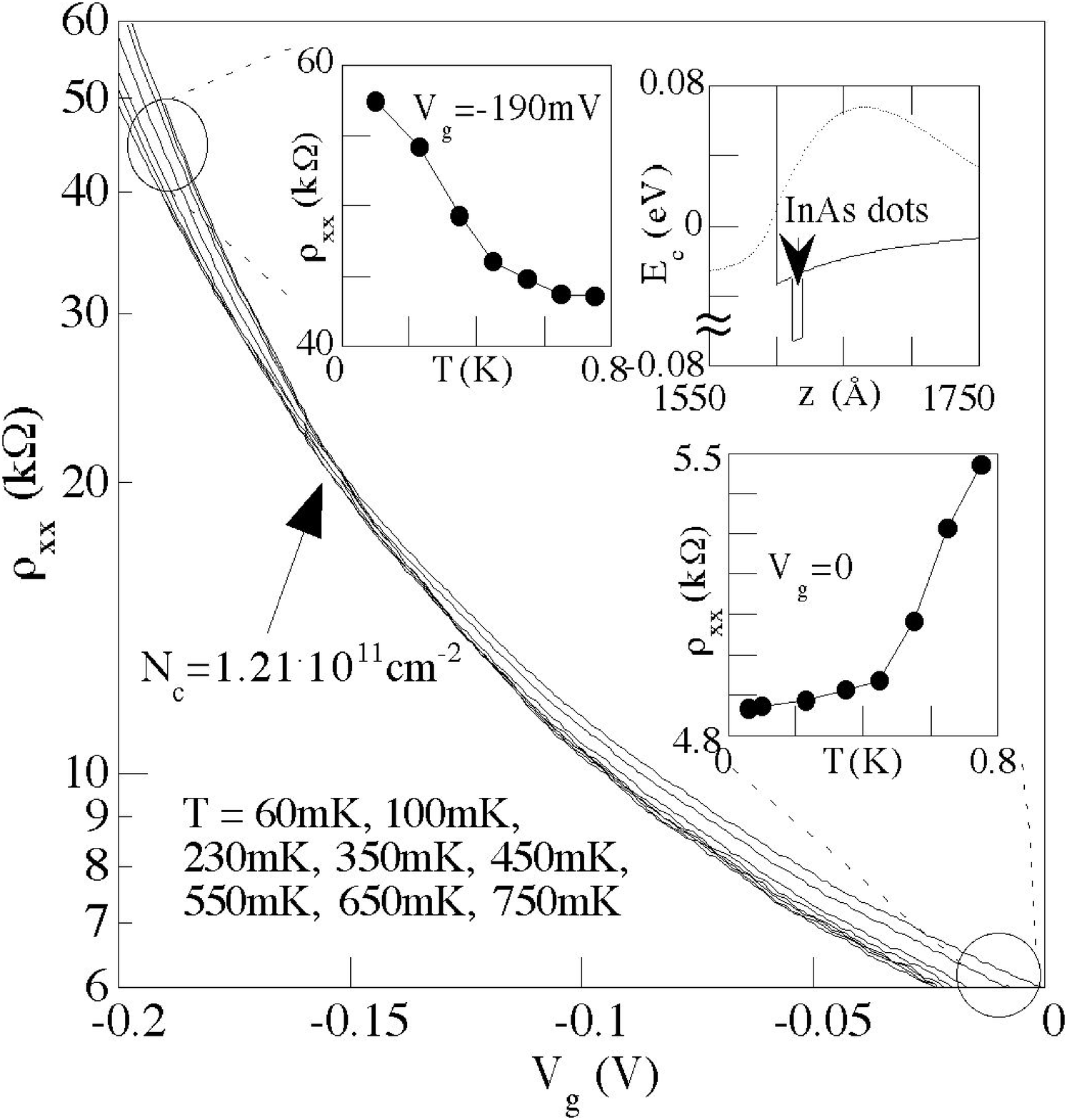}}
\caption{Resistivity as a function of gate voltage 
for a set of temperatures ranging from $60 mK$ to $750 mK$. Enlargements 
of the insulating and conducting regime are shown as insets (solid 
lines are guides to the eye).  Upper right inset: sketch of the band structure around the heterointerface, 
including the potential of a InAs self-assembled quantum dot, and the 
wave function of the lowest subband(dotted line).}
\end{figure}

In the following, we focus on the sample with the high dot density. Fig. 1 shows $\rho_{xx}$ traces 
as a function of gate voltage at zero 
magnetic field for temperatures between  $T = 60 mK$ and $750 mK$. 
The traces cross each other at a 
resistivity slightly below  $h/e^{2}$ and at a gate voltage of  $V_{g} \approx  -160mV$, 
corresponding to $N_{C}= 
1.21\cdot 10^{11} cm^{-2}$. At lower densities ($Vg<-160mV$), the sample 
behaves insulating, i.e. the resistivity 
increases with decreasing 
	temperature (upper left inset in Fig. 1). At densities above the 
fix point, the resistivity 
	drops with decreasing T, indicating a metallic character (lower right
inset in Fig. 1). The T dependence 
	saturates around $T=750mK$. The saturation of $\rho_{xx}$ in the metallic phase 
at low T is {\it not} due to  current heating, since the same traces do not saturate in the insulating 
regime. Rather, the T dependence 
	can be fitted by$\rho = \rho_{0}+\rho_{1}\cdot e^{-T_{0}/T}$, as observed in\onlinecite{Hanein98} and predicted 
theoretically from two different 
models\onlinecite{Song98},\cite{Pudalov97}. Since the 
fix point of the MIT is  poorly defined, scaling according to 
$\rho(T,N_{S})=f_{1}(|\delta _{N}|/T^{1/z\nu }),
(\delta _{N}= (N_{S}-N_{C})/N_{C}$, and $f_{1}$denotes the scaling function 
in T) is difficult to confirm, although from scaling attempts (not shown) 
we can estimate $z\cdot \nu = (2.6\pm 0.8)$. 
Samples with lower dot density from the same wafer did not display a 
MIT at B=0. They displayed 
insulating behavior for all available carrier densities, with the 
temperature dependence of the 
resistance becoming very small for high carrier densities. 
In Fig. 2a, $\rho_{xx}$  as a function of the electric 
field $E$ is shown for various electron densities. Again, a MIT is 
clearly visible, similar to the results 
in\onlinecite {Kravchenko96}. $N_{C}$ and $\rho_{xx}$ at the 
transition are the same as those 
found in the T dependence. The inset in Fig.2a shows the result of 
a similar experiment on a sample 
with lower dot density of about $4 \cdot 10^{10}$ dots $cm^{-2}.$ No metallic 
phase could be found. 
Fig. 2b shows the result of scaling according to  $\rho(E,N_{S})=f_{2}(|\delta  
_{N}|/E^{1/(z+1)\nu })$. For the best visual 
collapse of the curves onto one trace, we find 
	$(z+1) \cdot \nu = (4.5 \pm 0.3).$ The scaled curves show a stronger asymmetry 
than those obtained from 
	scaling in Si MOSFETs\onlinecite {Kravchenko96}, but similar to the asymmetry observed 
for hole systems in GaAs\onlinecite {Simmons98}. This asymmetry, however, is not surprising, since it is a 
measure for the nonlinearity of the 
	$\beta -$function around the critical point\onlinecite{Dobro97}. Each individual curve 
contains several points that 
	deviate from the scaled function. These points correspond to the low 
electric field range 
	$(E<20 mV / cm,$ Fig.2a). At these small electric fields, the 
	resistivity increases slightly and may indicate are-entrant behavior 
	into an insulating phase. Due to the small 
	sample size,  however, the experimental error is larger at small electric 
	fields. This low-electric field regime at high electron densities 
	needs further study and is not discussed here. 
	Combining the results for the scaling exponent obtained from 
electric field scaling and the 
	estimated value for the temperature scaling as described above, we 
find $z=(1.4\pm 1.0)$ and 
	$\nu=(1.9 \pm 0.9).$ Due to the large error bars, we do not attempt to 
	give a detailed interpretation.
	However, these numbers are in agreement with the values reported  
	in\onlinecite{Kravchenko96} and 
in agreement with theoretical 
	considerations, i.e. $z = 1$ for an interacting system \onlinecite{Sondhi97} and 
	$\nu = 4/3 $
from percolation theory\onlinecite{Song98}.
At present, we do not 
have a full understanding of the 
	existence of a metallic phase in our samples. It is clear, however, 
that a high density of SAQDs is needed for its formation. In the following, we speculate on possible 
explanations on what the metallic phase consists of and discuss
magnetoresistance measurements.	
Within a two-phase model\onlinecite{Song98},
	 a gaseous phase can coexist with a liquid phase 
which can undergo a percolation 
	threshold that defines the transition point from a metal to an 
insulator. In our sample, the 
	liquid phase could be formed by the disordered electron gas, while the 
localized gas phase may consist of minibands or quasi-bound states that 
emerge from the SAQDs. \\
 
\begin{figure}
\centerline{\epsfxsize=3.2in\epsfbox{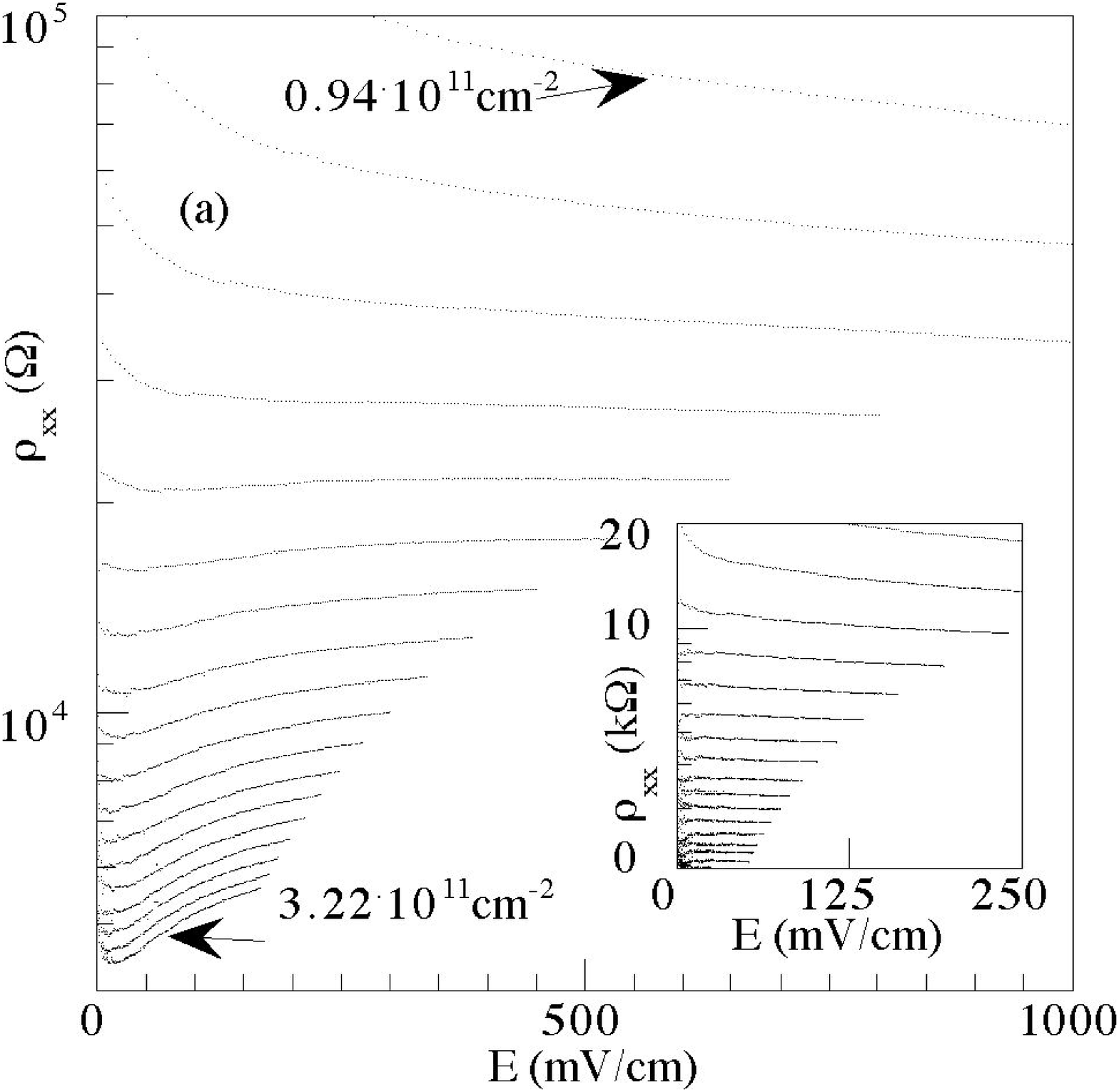}}
\centerline{\epsfxsize=3.2in\epsfbox{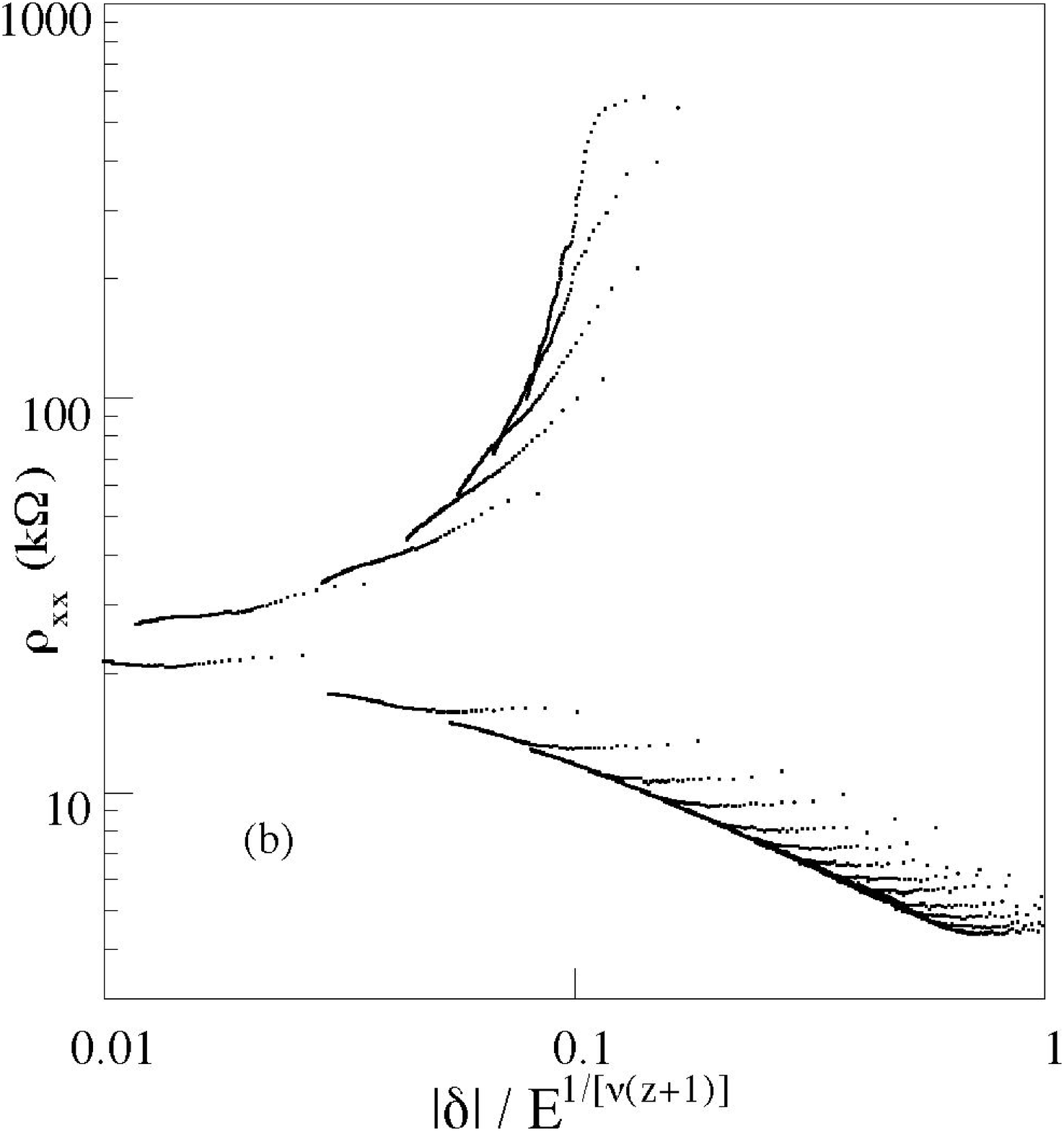}}
\caption{a:	Resistivity as a function of the electric 
field at carrier densities of 0.94 (uppermost curve), 1.05, 1.16 , 1.27, 1.38, 
1.50, 1.62, 1.75, 1.88, 2.01, 2.15, 2.29, 2.44, 2.59, 2.74, 2.89, 3.05, and 3.22 
(lowermost curve), in units of $10^{11}cm^{-2}$. The inset shows the resistivity 
of a sample with a lower density of SAQDs. The electron densities are the same as in 
the sample with the high dot density.
b: Scaling plot of the data of Fig. 2a. The scaling exponent $(z+1)\nu$
is ($4.5\pm 0.3)$.}
\end{figure}
The conduction band is lowered in regions were 
the InAs dots form. For the first electron in the empty conduction 
band this leads to an attractive potential well. As more electrons 
populate the conduction band they first fill the InAs dots and then 
build up a Fermi sea. For the mobile electrons in the GaAs conduction 
band the filled InAs quantum dots now represent repulsive scattering 
centers. The potential landscape can be viewed as a random antidot lattice 
with possible short-range ordering. From our previous experiments  we know 
that each dot is populated by about two electrons\onlinecite{Ribeiro98}. The depth of the 
InAs wells (100-200 meV) is about an order of magnitude larger than 
the Fermi energy ($\approx 5 meV$).  
However, the presence of the InAs dots 
can lead to quasi-bound states in the continuum of the dots close to 
the range of the Fermi energy. Since the InAs dots are rather 
homogeneous in size (the dot diameter varies by only about 7\% 
\cite{Leonhard93} ) 
and energetic structure\cite{Drexler94}, these quasi-bound states are expected to 
be rather narrow as well. In principle, mobile electrons can scatter 
into and out of such quasi-bound states which leads to a decreased 
conductivity compared to the 2DEGs without SAQDs. The states that are 
available for the electrons to scatter into are within an energy 
window of $k_{B}T$.
In this 
picture, a lower electron temperature would lead to a smaller number 
of available states and thus to an 
increasing conductivity with decreasing temperature, as in a metallic 
state. A similar situation is considered in\cite{Ulloa97}, where the 
structure of 2D electronic states in a strong magnetic field in the 
presence of a large number of resonant scatterers is calculated. So 
far, in transport experiments on electron gases with InAs SAQDs 
nearby\cite{Sakaki95}, such states have not been observed; however, note that 
higher dot densities than in these experiments are needed. In case a 
short-range order between the SAQDs exists, a local bandstructure may 
arise leading to an increased effective mass. This could 
give rise to an enhancement of the Coulomb interaction.
\begin{figure}
\centerline{\epsfxsize=3.2in\epsfbox{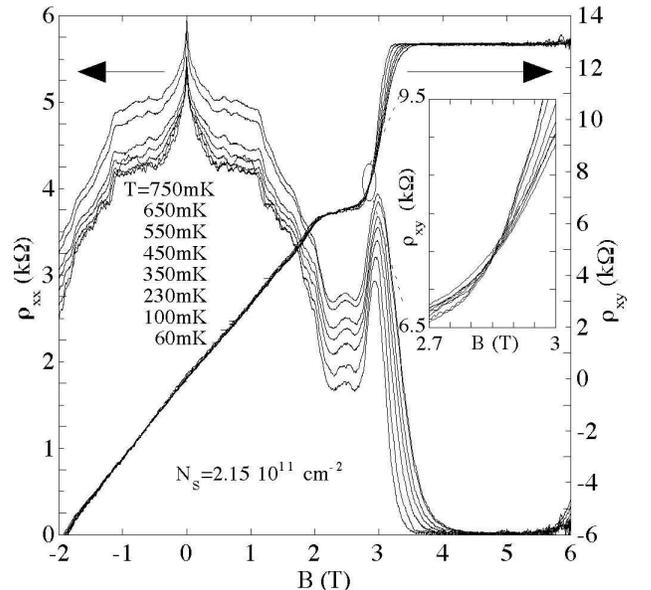}}
\caption{Longitudinal and Hall resistivity for various temperatures in the metallic regime. Since the electron 
gas is in the metallic state at B=0, no insulator-to quantum Hall 
liquid transition in $\rho_{xx}$ is observed. The Hall resistance, 
however, shows a fix point 
at $B=2.85T$ (inset).}
\end{figure}
     Clearly, the electrons form a quantum Hall liquid at sufficiently 
	high magnetic fields in all our 
samples. The longitudinal magnetoresistivity $\rho_{xx}$
	as well as the Hall resistivity $\rho_{xy}$ for the sample with the 
	high dot density shown in Fig.3. At
	 $B = 0$, the sample  is in a metallic 
	 state. Above $B = 4T$, the sample is in a quantum Hall state with 
filling factor 2 (i.e. 2 Landau levels are occupied). In $\rho_{xy}$, a fix point occurs at a magnetic field 
of $B=2.8 T$ 
as shown in the inset. Such 
	 magnetic field driven insulator-to-quantum Hall-liquid transitions 
have been experimentally 
	 observed in other disordered 2DEGs (for a review, see 
	 \onlinecite{Sondhi97} and references therein). In our system, a fix point occurs in 
$\rho_{xy}$, while $\rho_{xx}$ increases for increasing temperature over the 
entire magnetic field range. 
	 A scaling analysis around the fix point in $\rho_{xy}$ reveals a 
scaling exponent of $\kappa=(0.22\pm 0.02)$, in good agreement with exponents 
found in standard spin-degenerate quantum Hall 
systems \cite{Wei90}. These transitions can also be observed in samples 
with lower density of InAs dots,
	 where a fix point is also observed in $\rho_{xx}$.  
This is in tune with the fact 
	 that these samples always display insulating behavior at $B=0$ and 
crossing points at finite 
	 magnetic fields are therefore possible. 
	 
In our high dot density sample, the metallic phase is not destroyed by the application 
of parallel magnetic fields. Under parallel magnetic fields up to 
$B=8T$, the electric field scaling of the metallic phase remains 
unchanged.
	
To conclude, our experimental data clearly indicate the occurrence 
of a metal-insulator transition at zero magnetic field in a disordered two-dimensional 
electron gas in a Ga[Al]As heterostructure. Scaling theory has been applied to both 
the temperature dependence and the electric field dependence. While electric field 
scaling works well, temperature scaling is poor since the fix point in 
temperature is poorly defined. We have estimated the scaling exponents for our sample to $z 
= (1.4\pm1.0) $ and $\nu=(1.9 \pm 0.9)$, in agreement with the current framework of scaling 
theory and percolation theory. We speculate that it is the special kind of disorder 
potential in our samples that possibly modifies the ratio between the electron-electron 
interaction energy and the kinetic energy and thus drives this transition. 
In that sense the metal-insulator transition in two-dimensional systems remains a 
research topic that is far from being understood and requires more 
experimental as well as theoretical work.
		
We are grateful to T. Ihn, C. F. Zhang, T. M. Rice, and S. Ulloa for helpful discussions. 
This work was supported by the Schweizerische Nationalfonds and by 
QUEST. E. R. acknowledges 
financial support from Funda\c{c}\~ao 
de Amparo \`a Pesquisa do Estado de S\~ao Paulo.



\end{multicols}
\end{document}